\documentclass[10pt]{iopart}
\usepackage[utf8]{inputenc}
\usepackage{harvard}
\usepackage{todonotes}
\usepackage{aas_macros}
\usepackage{amssymb}

\begin{document}

\title{Introducing Physical Warp Drives}

\author{Alexey Bobrick, Gianni Martire}

\address{Advanced Propulsion Laboratory at Applied Physics, 477 Madison Avenue, New York, 10022, U.S.}
\ead{alexey.bobrick@appliedphysics.org}
\vspace{10pt}
\begin{indented}
\item[]September 2020
\end{indented}

\begin{abstract} 
The Alcubierre warp drive is an exotic solution in general relativity. It allows for superluminal travel at the cost of enormous amounts of matter with negative mass density. For this reason, the Alcubierre warp drive has been widely considered unphysical. In this study, we develop a model of a general warp drive spacetime in classical relativity that encloses all existing warp drive definitions and allows for new metrics without the most serious issues present in the Alcubierre solution. We present the first general model for subluminal positive-energy, spherically symmetric warp drives; construct superluminal warp-drive solutions which satisfy quantum inequalities; provide optimizations for the Alcubierre metric that decrease the negative energy requirements by two orders of magnitude; and introduce a warp drive spacetime in which space capacity and the rate of time can be chosen in a controlled manner. Conceptually, we demonstrate that any warp drive, including the Alcubierre drive, is a shell of regular or exotic material moving inertially with a certain velocity. Therefore, any warp drive requires propulsion. We show that a class of {\it subluminal, spherically symmetric} warp drive spacetimes, at least in principle, can be constructed based on the physical principles known to humanity today.

\end{abstract}

%

%
\submitto{\CQG}
%
%
%

\section{Introduction}
\label{sec:Intro}


The classical-relativistic Alcubierre drive solution allows timelike observers to travel superluminally, although at the expense of using material with negative rest-mass energy \cite{Alcubierre1994}; for recent reviews see also \cite{Lobo2007,Alcubierre2017}. This solution is given by the following asymptotically-flat metric:
\begin{equation}
    {\rm d}s^2=-c^2 {\rm d} t^2 + ({\rm d}x-f(r_s)v_s {\rm d}t)^2+{\rm d}y^2 + {\rm d}z^2,
    \label{eq:Alc}
\end{equation}
where $r_s=\sqrt{(x-v_st)^2 + y^2 + z^2}$. The metric describes a spherical warp bubble (a region deviating from the flat metric) moving along the $x$-axis with an arbitrary velocity $v_s$, which may be larger or smaller than the speed of light $c$. 

The shape function $f(r_s)$, present in the metric, defines the size and profile of the warp bubble. For large distances $r_s$ from the bubble, $f(r_s)=0$ and the spacetime is flat. For small distances, $r_s\approx0$, the shape function $f(r_s) = 1$, and the metric describes the flat internal region of the bubble. In a coordinate system with $x\rightarrow x^\prime = x-v_st$, this internal region is described explicitly by the Minkowski flat metric. The intermediate region, for which $f(r_s)\gtrsim 0$, corresponds to the spherical boundary of the warp. In the original study, \cite{Alcubierre1994} chose function $f(r_s)$, somewhat arbitrarily, as:
\begin{equation}
\label{eq:ShFun0}
f_{\rm Alc} (r_s) = \frac{\tanh (\sigma (
r_s + R )) - \tanh (\sigma ( r_s - R ))}{2 \tanh( \sigma R)},
\end{equation}
where parameters $R$ and $\sigma^{-1}$ define the radius and the thickness of the transition from the internal to the external region, correspondingly.

In the case of superluminal motion, the metric possesses a black hole-like event horizon behind the bubble and a white hole-like event horizon in front of it \cite{Finazzi2009}. These event horizons arise because timelike observers cannot exit the superluminal ship in the direction ahead of it, and cannot enter it from behind. In both cases, the timelike observers would have to move superluminally when outside of the ship.

The energy density for the Alcubierre drive, as measured by Eulerian observers ($u_\mu = (1,0,0,0)$), is given by:
\begin{equation}
\label{eq:AlcT}
    T^{00}=-\frac{1}{8\pi}\frac{\rho^2 v_s^2}{4r_s^2}\left(\frac{{\rm d} f}{{\rm d} r_s}\right)^2,
\end{equation}
where $\rho^2 \equiv y^2 + z^2$ is the cylindrical coordinate.


Despite its interesting properties which allow timelike observers to travel at arbitrary velocities, the Alcubierre drive solution possesses several drawbacks. As noted earlier, it requires negative energy densities, Equation~(\ref{eq:AlcT}), and thus violates the weak energy condition. Although negative energy densities are a general property of any superluminal drive \cite{Olum1998,Visser2000},  the energy density is also negative at subluminal speeds for the Alcubierre drive, even in the weak-field approximation \cite{Lobo2004}. Additionally, superluminal motion allows for closed timelike loops, e.g. leading to grandfather paradox, and violates the null energy condition and causality, e.g. \cite{Everett1996}, although the latter may be recovered at the expense of Lorentz invariance \cite{Liberati2002}. When moving superluminally, the drive has an additional problem. It leads to quantum instabilities related to pair production near the horizon behind the warp as well as accumulation of particles at the horizon in the front part of the warp \cite{Finazzi2009}.

The Alcubierre drive is also problematic at sub-relativistic speeds. Firstly, it requires unphysically large amounts of (negative) energy. For instance, it would require an amount of negative energy comparable to the mass of the Sun to produce relativistic bubbles of $\approx \rm{meter}$ sizes \cite{Alcubierre1994}. Furthermore, such high {\it negative} energy densities do not appear even theoretically feasible. There are no known materials which would allow for gathering large amounts of negative energy in a controlled way. While zero-point vacuum fluctuations may produce negative energies in curved spacetimes, for Alcubierre drives this situation is only possible if the walls of the bubble had thicknesses comparable to Planck scales. Such thin walls, however, require extreme amounts of energy -- comparable to the rest-mass energy of the Universe -- as may be seen from Equation~(\ref{eq:AlcT}). Therefore, there is no physical way to create an Alcubierre drive \cite{Pfenning1997CQG,Ford1997}.

Finally, there is no proposed way of creating an Alcubierre drive, even if negative energy were available. In the original study, \cite{Alcubierre1994} suggested that the velocity may be time-dependent, $v_s = v_s(t)$. Indeed, Equation~(\ref{eq:AlcT}) retains its form even for time-dependent velocities. And, since $v_s=0$ corresponds to flat spacetime, it was assumed that the Alcubierre drive might be generated through acceleration. However, the metric in Equation~(\ref{eq:Alc}) with time-variable $v_s = v_s(t)$ corresponds to a time-variable stress-energy tensor which does not satisfy continuity equations. Alternatively, such solutions may be said to require an implicit dynamical field to effectively provide propulsion for the object, e.g. \cite{Bassett2000}. Generally, there are no self-consistent warp drive solutions proposed in the literature which can self-accelerate at all from zero velocities, not to mention gain superluminal speeds.


Despite the rather extensive work on the properties of the Alcubierre drive solution, it remains unclear which of the above issues are features of the Alcubierre solution specifically or more fundamental properties of warp drives as such. New warp drive solutions have been introduced only in very few studies. \cite{vandenBroeck1999} reduced the energy requirements of the Alcubierre drive to about the mass of the Sun while satisfying the vacuum energy inequalities. The reduction was realized by decreasing the externally measured size of the warp bubble down to $10^{-15}\,{\rm m}$ while keeping the internal volume constant. This solution satisfies the weak energy conditions, although it requires that classical gravity remains applicable down to such small scales, at which it was never tested. However, as we show in \ref{sec:AppA} through a coordinate transformation, this solution is equivalent to the Alcubierre solution.

\cite{Natario2002} constructed a warp drive solution without space contraction or expansion, contrary to the earlier assumption that it facilitated the movement of warp drives. \cite{Natario2006} constructed a new subluminal warp drive solution in the weak-field regime, which required negative energies. \cite{Loup2001} had previously introduced a modified version of the Alcubierre drive intended to alter the rate of time for the observers inside the bubble. However, their modification reduces to the original Alcubierre metric, as we also show in \ref{sec:AppA}. Finally, \cite{Lentz2020} has recently proposed a warp drive metric claiming to have purely positive energy everywhere in both subluminal and superluminal regimes, although without providing means to reproduce the study.

The works above, to our knowledge, summarize all the modifications of the Alcubierre drive available in the literature. Superluminal travel had also been studied by \cite{Krasnikov1998} and \cite{Everett1997}. In these studies, the authors introduced Krasnikov tubes. Krasnikov tubes are `spacetime tunnels' which allow for superluminal travel without violating causality, but only for round trips and with much larger energy requirements than the Alcubierre drive. Superluminal travel has also been discussed in the context of wormholes, e.g. \cite{Garattini2007}, and time-machine metrics, e.g. \cite{Fermi2018}, in all cases requiring negative energies. Finally, modified gravity theories may provide some desirable properties for the Alcubierre drive. For instance, conformal gravity allows for construction of Alcubierre solutions with positive energy only \cite{Varieschi2013}, while extra-dimensional theories of gravity may reduce the energy requirements of the drive \cite{White2013}.

In this study, we show that the properties of the Alcubierre metric -- in particular, its negative energy density and the accompanying immense energy requirements -- are not a necessary feature of warp drive spacetimes. In Section~2, we discuss that any general warp drive, including the Alcubierre metric, may be thought of as a shell of positive- or negative-energy density material which modifies the state of spacetime in the flat vacuum region inside it. In Section~3, we introduce, for the first time, the most general spherically symmetric warp drives. We show that the reason for the negative energy requirements of the Alcubierre metric and all the warp drives introduced in the literature is, likely, the truncation of the gravitational field outside of the metric, and the accelerated rate of time compared to the comoving free observer. In turn, the most general positive-energy (spherical) warp drive solutions must slow down the time compared to the comoving observer and have a gradual (Schwarzschild) fall off of the gravitational field. In Section~4, we introduce a simple method to construct warp drive metrics by choosing the state of spacetime inside them. Additionally, we show that the Alcubierre metric is only a particular member of a much more general class of warp drive solutions admitting negative energy densities. Subsequently, we construct several spacetimes with a range of new properties. We introduce a macroscopic negative-energy drive which satisfies quantum inequalities, as well as several metrics with negative-energies which allow one to control the rate of time and spatial contraction inside the drive. In Section~5, we review the interpretation of warp drives in light of this study.

\section{General class of warp drives}
\label{sec:GenDrives}
\subsection{Definition of a general warp drive spacetime}

\begin{figure}
\begin{center}
    \includegraphics[width=\linewidth]{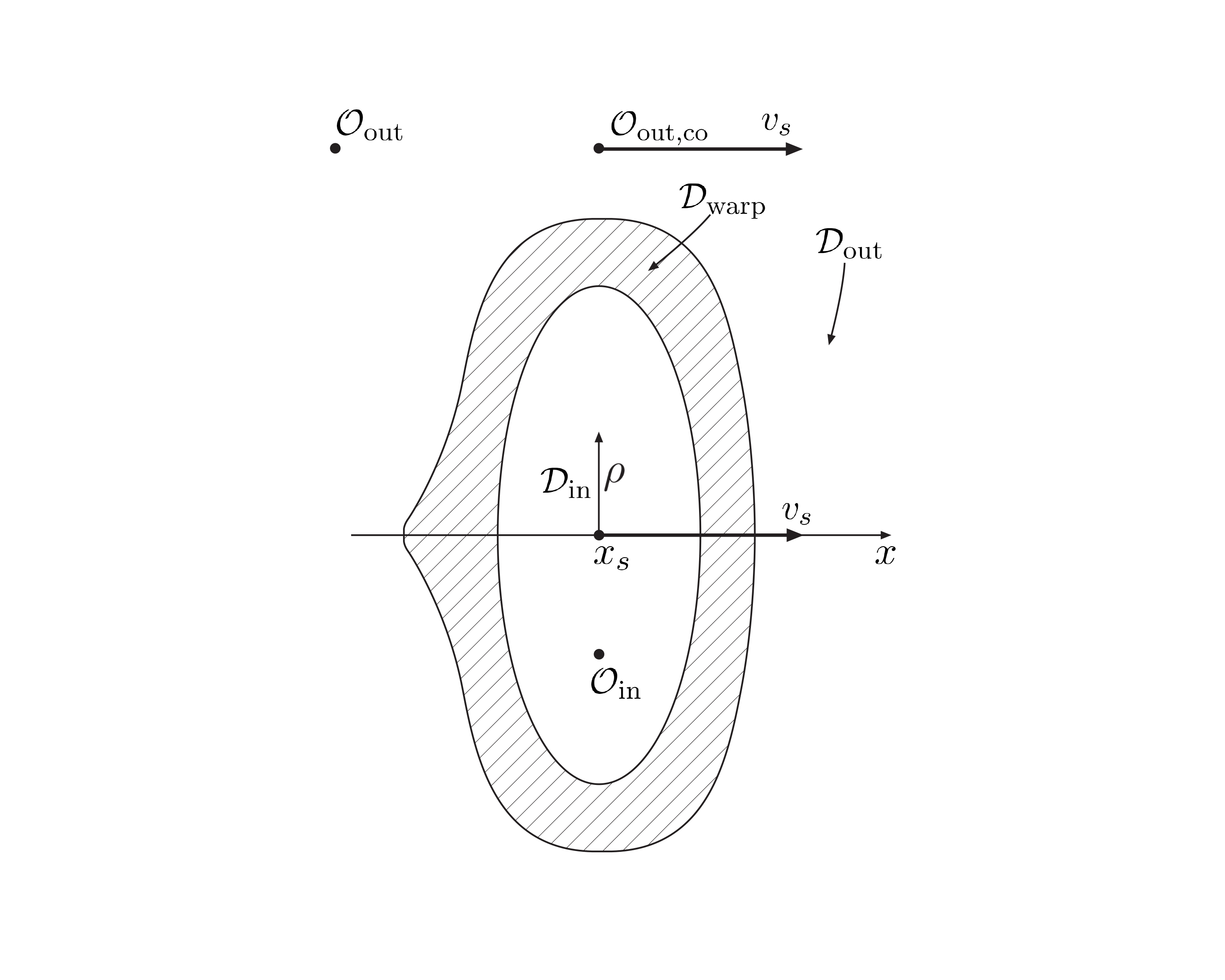}
\end{center}
\begin{flushleft}
    \caption{A schematic illustration of a warp-drive spacetime. The spacetime consists of three regions: Asymptotically-flat vacuum background $\mathcal{D}_{\rm out}$ (background), general stationary curved region $\mathcal{D}_{\rm warp}$ with a spherical topology (the warping region) and a flat inner region $\mathcal{D}_{\rm in}$ (`passenger' space). Any such spacetime, including the Alcubierre drive metric, is realised through a shell of ordinary or exotic negative energy density material filling the warping region $\mathcal{D}_{\rm warp}$. Axis $x$ shows the direction of motion, while $\rho$ is the cylindrical radius. As we discuss in Section~\ref{sec:AxiClasses}, flattened disk-shaped metrics minimise energy requirements of the particular Alcubierre, but not necessarily other, warp drive spacetimes. As we also discuss in Section~\ref{sec:Disc}, warp drive spacetimes require some form of propulsion in order to accelerate. For this reason, in physical realisations of such spacetimes, the front and rear parts are likely asymmetric. \label{fig:DriveScheme}}
\end{flushleft}    
\end{figure}

We define a general warp drive spacetime, as shown in Figure~\ref{fig:DriveScheme}, as an asymptotically-flat vacuum region $\mathcal{D}_{\rm out}$ (background) which encloses a compact arbitrarily curved region $\mathcal{D}_{\rm warp}$ with a spherical topology (the warping region); $\mathcal{D}_{\rm warp}$, in turn, encloses a flat extended\footnote{To avoid effectively including general asymptotically-flat spacetimes in the defintion, we require that the extent of the region $\mathcal{D}_{\rm in}$ should be non-vanishing and physically interesting. In other words, the inner region $\mathcal{D}_{\rm in}$ should be large enough to allow observers to conduct physical experiments of desired scales.} compact region $\mathcal{D}_{\rm in}$ with a trivial topology (the `passenger' area). This definition covers both all the existing warp drive spacetimes, such as the one by \cite{Alcubierre1994}, as well as trivial solutions, such as weak-field perturbations of the Minkowski spacetime. To formalize the difference between extreme and mild cases like these, we shall consider the warped region $\mathcal{D}_{\rm warp}$ by using comoving reference frames located inside and outside it, which will allow us to quantify how strong its gravitational influence is.

We shall focus on warp drive spacetimes which, intuitively, are stationary or non-changing from the `passenger' point of view. More formally, we consider warp drive spacetimes which{, by definition,} admit a global Killing vector field, ${\bf \xi}$, which is aligned with the four-velocity of the boundary of the region $\partial\mathcal{D}_{\rm in}$ (the inner boundary of the shaded region in Figure~\ref{fig:DriveScheme}). Such a field establishes the global frame of rest with respect to the warped region. Subsequently, whenever we discuss a physical motion of an observer relative to the warp drive, we consider, locally, the motion relative to the global reference frame defined by the field ${\bf \xi}$. Furthermore, using this vector field will allow us to apply the well-known techniques suitable for spacetimes possessing a timelike Killing vector to analyse and classify warp drive spacetimes.

Since the internal region $\mathcal{D}_{\rm in}$ is flat, vector field ${\bf \xi}$ is constant in that region. Therefore, in the internal region, this vector field ${\bf \xi}$ defines a reference frame (tetrad) for the internal observer\footnote{While, in relativity, physical observers are timelike by definition, we shall occasionally be mentioning timelike and spacelike observers for convenience. By these, we shall understand physical observers and spacelike frames, correspondingly.}  $\mathcal{O}_{\rm in}$. Similarly, in the asymptotic infinity, vector field ${\bf \xi}$ defines the frame of the remote comoving observer $\mathcal{O}_{\rm out, co}$. This comoving reference frame $\mathcal{O}_{\rm out, co}$, as we further argue, is the most natural frame against which one can compare the frame $\mathcal{O}_{\rm in}$ inside the warped region. Finally, this remote comoving observer $\mathcal{O}_{\rm out, co}$ may be moving with constant three-velocity ${\bf v}_s$ with respect to a timelike observer $\mathcal{O}_{\rm out}$, the latter of which we shall consider to be at rest. Everywhere in this study, apart from the general discussion in this section, we shall consider warp drives spacetimes which are axisymmetric along the direction of motion.

The three-velocity of the comoving observer ${\bf v}_s$, { which represents the velocity of the warped region relative to the remote observer $\mathcal{O}_{\rm out}$}, may be slower or faster than the speed of light with respect to the observer. This corresponds to four-velocity of the comoving observer $\mathcal{O}_{\rm out, co}$ being timelike, null or spacelike. Both observers $\mathcal{O}_{\rm in}$ and $\mathcal{O}_{\rm out, co}$, therefore, may in general, at least formally, be timelike, null, or spacelike. If region $\mathcal{D}_{\rm warp}$ contained vanishing stress-energy tensor, the whole spacetime would be close to Minkowski spacetime and the two observers $\mathcal{O}_{\rm in}$ and $\mathcal{O}_{\rm out}$ would both be timelike, null, or spacelike at the same time. However, it is also possible for region $\mathcal{D}_{\rm warp}$ to be sufficiently curved such that, as we show further, the norms of the four-velocities of observers $\mathcal{O}_{\rm in}$ and $\mathcal{O}_{\rm out,co}$ may be different from each other. Therefore, all stationary warp drive spacetimes may be split into four distinct classes, based on whether the remote comoving observer  $\mathcal{O}_{\rm out,co}$  moves subluminally (subluminal drives) or superluminally (superluminal drives), and whether the internal observer $\mathcal{O}_{\rm in}$ has the same norm as the comoving observer  $\mathcal{O}_{\rm out,co}$ (mild warp drives) or different norm from the comoving observer (extreme warp drives):
\begin{description}
    \item[Class I: Mild subluminal warp drives:] These spacetimes are defined by the vector field ${\bf \xi}$ being timelike everywhere. Consequently, three-velocities of such drives are subluminal, i.e. $v_s<c$. Spacetimes of this class approach the flat Minkowski spacetime in the trivial limit, and observers $\mathcal{O}_{\rm in}$, $\mathcal{O}_{\rm out, co}$ reduce to a pair of co-moving timelike Lorentz observers. Non-trivial members of this class contain spacetimes with region $\mathcal{D}_{\rm warp}$ sufficiently curved, so that tetrads of observers $\mathcal{O}_{\rm in}$ and  $\mathcal{O}_{\rm out, co}$ differ significantly from each other, i.e. the observers read off different rates of clocks and lengths of rulers. At the same time, such spacetimes also contain weak-field solutions corresponding to classical shell-like objects moving with subluminal velocities and weakly modifying the state of the spacetime inside them. Such solutions are possible because the $\mathcal{D}_{\rm warp}$ region may be set arbitrarily close to being flat, rendering the whole spacetime arbitrarily close to Minkowski spacetime.
    
    \item[Class II: Mild superluminal warp drives:] These spacetimes are characterized by the vector field ${\bf \xi}$ being spacelike or null everywhere. Consequently, such warp drives have luminal or superluminal velocities, i.e., $v_s\geq c$. These spacetimes also admit a trivial limit, wherein they reduce to flat Minkowski spacetime, with the tetrads $\mathcal{O}_{\rm in}$, $\mathcal{O}_{\rm out, co}$ corresponding to a pair of comoving null or spacelike (superluminal) `observers'. Weak-field members of the class correspond to small amounts of `superluminal matter' in the region $\mathcal{D}_{\rm warp}$ introducing small differences in the measurements of frames $\mathcal{O}_{\rm in}$, $\mathcal{O}_{\rm out, co}$. By `superluminal matter' we understand the matter at rest with respect to a space-like reference frame. In the case of the stress-energy tensor for a perfect fluid, such matter violates the dominant energy condition. A general spacetime of this class introduces non-trivial differences between frames $\mathcal{O}_{\rm in}$ and $\mathcal{O}_{\rm out, co}$. Since superluminal matter cannot be produced from physical matter, and since null or spacelike tetrads cannot be associated with physical observers, the spacetimes of this class have limited interest.
    
    \item[Class III: Extreme superluminal warp drives:] These spacetimes are defined by the vector field ${\bf \xi}$ being timelike in the inner region $\mathcal{D}_{\rm in}$, but null or spacelike in the asymptotic infinity of the outer region $\mathcal{D}_{\rm out}$. The remote comoving observers in such spacetimes move luminally or superluminally relative to the resting timelike observer $\mathcal{O}_{\rm out}$, i.e., $v_s\geq c$. This class of spacetimes does not contain trivial solutions and the warped region $\mathcal{D}_{\rm warp}$ is sufficiently curved to allow timelike observers $\mathcal{O}_{\rm in}$ to be moving superluminally relative to the timelike observer $\mathcal{O}_{\rm out}$ (and as a consequence, the timelike observer $\mathcal{O}_{\rm in}$ will be travelling back in time from the point of view of yet another remote timelike observer $\mathcal{O}_{\rm out}^\prime$). At the same time, the comoving observer $\mathcal{O}_{\rm out, co}$ is formally superluminal; in other words, remote timelike observers cannot be comoving with a warp drive of this class. For such spacetimes, we can define a Killing horizon $\partial\mathcal{D}_{\rm K}$ as the minimal boundary surface where the Killing vector field ${\bf \xi}$  becomes null. A necessary, but not sufficient, condition for warp drives of this class to be physical is that the Killing horizon $\partial\mathcal{D}_{\rm K}$ does not intersect with region $\mathcal{D}_{\rm warp}$. If the Killing horizon did intersect with region $\mathcal{D}_{\rm warp}$, some part of the matter in that region would be at rest with respect to spacelike tetrads aligned with vector field ${\bf \xi}$. In other words, a fraction of matter in the region $\mathcal{D}_{\rm warp}$ would have to be superluminal (in the same sense, as for Class II above) and violate the dominant energy condition.
    
    \item[Class IV: `Extreme' subluminal warp drives:] Spacetimes of this class are defined by the vector field  ${\bf \xi}$ being null or spacelike in the inner region $\mathcal{D}_{\rm in}$, but timelike at the asymptotic infinity of the outer region $\mathcal{D}_{\rm out}$. Since the comoving observer $\mathcal{O}_{\rm out, co}$ is timelike, such spacetimes are subluminal, i.e. $v_s < c$. Since the Killing vector is spacelike in the inner region $\mathcal{D}_{\rm in}$, no timelike internal observers can be at rest relative to the inner boundary of the drive $\partial\mathcal{D}_{\rm in}$. This property bears similarity to black hole spacetimes, except for that the inner region $\mathcal{D}_{\rm in}$ for this class of spacetimes is flat everywhere. A necessary condition for spacetimes of this class to be physical is that the Killing horizon coincides with the boundary of the inner region, $\partial\mathcal{D}_{\rm in}$. Otherwise, some fraction of matter in the region $\mathcal{D}_{\rm warp}$ will be superluminal, as in the previous class.
\end{description}

In practical terms, as follows from our definition, any warp drive spacetime may be seen as a shell of ordinary or exotic material which fills region $\mathcal{D}_{\rm warp}$ and moves with some constant velocity relative to an external timelike observer $\mathcal{O}_{\rm out}$. The presence of the shell inevitably modifies the space and time in the inner region $\mathcal{D}_{\rm in}$. This modification leads (in addition to other effects) to a difference in the measurements of times and lengths between an observer in the inner region $\mathcal{D}_{\rm in}$ and a comoving remote observer $\mathcal{O}_{\rm out, co}$. General warp drive spacetimes, therefore, form a continuous family which includes both trivial (flat or nearly-flat) and non-trivial (strongly curved) spacetimes.

\subsection{How do existing warp drive spacetimes fit in the classification?}
\label{sec:ClassReview}
The Alcubierre drive is a truncated warp drive spacetime defined by meeting the three requirements. First, that the external spacetime $\mathcal{D}_{\rm out}$ is Minkowskian. In other words, the matter in the warped region $\mathcal{D}_{\rm warp}$ does not exhibit gravitational influence outside of it (Requirement 1). Second, observers $\mathcal{O}_{\rm in}$ and $\mathcal{O}_{\rm out}$ are always timelike, and the tetrads of these observers are equal to each other. In other words, the clock rates and the lengths of the rulers of observers $\mathcal{O}_{\rm in}$ and $\mathcal{O}_{\rm out}$ are synchronized and put equal to each other (Requirement 2). Finally, the warped region $\mathcal{D}_{\rm warp}$ is functionally limited to Equations~(\ref{eq:Alc}) and (\ref{eq:ShFun0}) in Section~\ref{sec:Intro} (Requirement 3). For subluminal velocities, the Alcubierre drive belongs to Class I, mild subluminal warp drives. For superluminal velocities it belongs to Class III, extreme superluminal warp drives.

Compared to general warp drives, this set of constraints may seem artificial. For example, requiring that the material of the Alcubierre drive does not gravitate in region $\mathcal{D}_{\rm out}$, which is non-typical for massive objects, as we discuss further in Section~\ref{sec:SphGen}, unnecessarily puts strong constraints on the stress-energy tensor in region $\mathcal{D}_{\rm warp}$. Similarly, requiring that the time of observers $\mathcal{O}_{\rm in}$ is synchronized with the time of observers $\mathcal{O}_{\rm out}$ means that, at least in the subluminal regime, the time of internal observers $\mathcal{O}_{\rm in}$ is accelerated compared to the time of remote comoving observers $\mathcal{O}_{\rm out, co}$. In comparison, general warp drive spacetimes do not need to be truncated and may allow an arbitrary relation between the measurements of observers $\mathcal{O}_{\rm in}$ and $\mathcal{O}_{\rm out, co}$. 

A more general class of warp drive metric was proposed by \cite{Natario2002}. However, their most general warp drive, similarly to the Alcubierre solution, still imposes Requirements 1 and 2 from the Alcubierre metric. That is, is also makes region $\mathcal{D}_{\rm out}$ strictly flat outside of the drive and synchronizes the time of the internal and the resting (not comoving) observers. Similarly to the Alcubierre solution, the \cite{Natario2002} class of spacetimes also belongs to Class I in the subluminal regime and Class III in the superluminal regime.

Further, in Sections~\ref{sec:Axi} and \ref{sec:SphGen}, we show that general warp drive spacetimes are diverse and can have a variety of properties that are often more appealing compared to either the Alcubierre spacetime or to the \cite{Natario2002} class of spacetimes.

\subsection{Properties of warp drive spacetimes}
\label{sec:DriveProps}

In this section, we summarise the standard tools applicable for analyzing general warp drives, in order to apply them further in Sections~\ref{sec:Axi} and \ref{sec:SphGen}.

The internal physical volume of a warp drive measured by observer $\mathcal{O}_{\rm in}$, inside the warped region, is given by $V_{\rm in}\equiv\int_{\mathcal{D}_{\rm in}}{\rm d}^3x^i_{\rm in}$. For truncated warp drive spacetimes, i.e. spacetimes where the outside vacuum region $\mathcal{D}_{\rm out}$ is flat, the external volume may be defined as $V_{\rm out}\equiv\int_{\mathcal{D}_{\rm in \cup warp}}{\rm d}^3x^i_{\rm out}$. The integral is calculated over the volume enclosing the warped region and assuming, during the integration, that the volume's interior is flat. Such a construction provides a measure of the size of the warp bubble as observed from outside. It may be formally done with the help of the Cartesian coordinate grid of the remote observer $O_{\rm out}$, analytically extended to cover the warped region. The comoving external volume $V_{\rm out, co}$ may be similarly defined for the comoving observer. For general asymptotically-flat metrics, we do not consider external volumes as they are coordinate-dependent.

We calculate the energy density, the momentum flux, and their volume integrals for warp drive spacetimes, and the mechanical stress distributions, in the coordinate system adapted to the remote comoving observer ${\mathcal{O}}_{\rm out, co}$. The coordinate system adapted to the comoving observer is aligned with the Killing vector field ${\bf \xi}$, which allows for a $3+1$ decomposition of the spacetime and thus provides coordinate-invariant measures for the total energy, mass and momentum of the drive. We also consider a system of Eulerian observers defined through four-velocities $u_{\mu}=\frac{1}{\sqrt{-g^{00}}}(1,0,0,0)$, similar to how it was done in \cite{Alcubierre1994}. These observers are imperatively timelike and are normal to the surfaces of constant time, but are coordinate-dependent.

The energy density $w$ in each of the coordinate systems is given by $w=(-g^{00})^{-1}T^{00}$. The momentum flux $j^i$ and the pressure $P$ in the same reference system are given by $j^i=u_{\mu} T^{\mu i}=\frac{1}{\sqrt{-g^{00}}} T^{0i}$ and $P=T^{\mu\nu}(\delta_{\mu\nu}-u_{\mu}u_{\nu})= T + (g^{00})^{-1}T^{00}$, where $T\equiv T^{\mu\nu}g_{\mu\nu}$. Notably, the total energy and momentum contain the contribution only from the warped region $\mathcal{D}_{\rm warp}$: $E=\int_{\mathcal{D}_{\rm warp}}(-g^{00})^{-1}T^{00}\sqrt{-g}{\rm d}^3 x^i_{\rm out}$, $J^i=\int_{\mathcal{D}_{\rm warp}}\frac{1}{\sqrt{-g^{00}}} T^{0i}\sqrt{-g}{\rm d}^3 x^i_{\rm out}$.

Warp drive solutions may be compared based on how they satisfy different energy conditions. Among these are the strong and weak energy conditions. Furthermore, we also consider quantum inequalities, e.g. \cite{Pfenning1997CQG}. These inequalities, at least in the weak-field regime, potentially allow the spacetime curvature to modify the quantum vacuum and, this way, allow for some negative-energy density:
\begin{equation}
\label{eq:EnCond}
    \frac{\tau_0}{\pi}\int_{-\infty}^{\infty}\frac{\langle T_{\mu\nu}u^\mu u^\nu\rangle}{\tau^2+\tau_0^2}{\rm d}\tau\ge -\frac{3}{32\pi^2\tau_0^4},
\end{equation}
where $\tau$ is the proper time of the observer and $\tau_0$ is an arbitrary constant smaller than the local radius of curvature.

\section{General spherically symmetric warp drives}
\label{sec:SphGen}

\subsection{General subluminal spherically symmetric drives}

One may gain considerable intuition into the nature of warp drive spacetimes by considering spacetimes which are spherically symmetric in the comoving reference frame. Since spherically symmetric spacetimes are fully solvable in relativity, we may achieve full characterization of such warp drive spacetimes. More formally, in this section, we limit ourselves to spacetimes which contain an ${\rm SO}(3)$ group whose orbits are orthogonal to the Killing vector field ${\bf \xi}$ introduced in Section~\ref{sec:GenDrives}. The Killing vector field ${\bf \xi}$ defines the local frame of rest in the comoving reference frame. As we comment further, the presence of such a group is possible only for class I and IV spacetimes, i.e., subluminal warp drives. 

Class I of subluminal warp drives -- mild subluminal warp drives -- contains trivial solutions and may be associated with ordinary objects, such as a thin shell of non-exotic material. However, the well-known Alcubierre and Natário drives, in the subluminal regime, also belong to the same class. By considering general spherically symmetric subluminal warp drives, we shall gain considerable intuition about these spacetimes and provide a possible explanation for their negative energy requirement.

The most general metric for a spherically symmetric stationary spacetime can be written in Schwarzschild coordinates as:
\begin{equation}
    {\rm d}s^2 = - N(r) c^2{\rm d}t^2+\Lambda(r){\rm d}r^2+r^2({\rm d}\theta^2+\sin^2\theta{\rm d}\varphi^2),
\label{eq:SphMetric}
\end{equation}
where functions $N(r)$ and $\Lambda(r)$ are general. Therefore, energy density of a spherically symmetric warp drive spacetime, in the comoving reference frame, is given by:
\begin{equation}
    w(r)=\frac{1}{8\pi r^2}\left(1-(\frac{r}{\Lambda(r)})^\prime\right)
    \label{eq:Sph}
\end{equation}
A necessary condition for the spherically symmetric metric in Equation~(\ref{eq:SphMetric}) (above) to describe a warp drive spacetime is that the energy density $w$ in both the inner region $\mathcal{D}_{\rm in}$ and outer region $\mathcal{D}_{\rm out}$ be equal to zero. In particular, as follows from Equation~(\ref{eq:Sph}), for any spherically symmetric warp drive, $\Lambda(r)=1$ in region $\mathcal{D}_{\rm in}$. Furthermore, integrating Equation~(\ref{eq:Sph}), we can express the metric function $\Lambda(r)$ directly through the energy density distribution:
\begin{equation}
    \Lambda(r)=\frac{1}{1-\frac{2}{r}\int_0^r 4\pi w(r^\prime) r^{\prime 2}{\rm d}r^{\prime}}
\label{eq:Sph2}    
\end{equation}

Equations~(\ref{eq:Sph}) and (\ref{eq:Sph2}) allow us to analyze the energy density distribution of spherically symmetric solutions and to construct such spacetimes from any desired energy distribution. Thus, we may see that for purely positive energy spacetimes, $\Lambda(r)>1$ in the outside region $\mathcal{D}_{\rm out}$. This result is in agreement with the Birkhoff's theorem, and states that the metric in the region $\mathcal{D}_{\rm out}$ is described by the Schwarzschild solution. In comparison, for truncated warp drive spacetimes, such as the Alcubierre or Natário solutions, the metric in the external region is assumed to be strictly flat. This assumption necessarily requires that the integral $\int_0^{r_{\rm out}} 4\pi w(r^\prime) r^{\prime 2}{\rm d}r^{\prime}=0$. As a result, any non-trivial truncated spherically symmetric warp drive must contain regions where the energy density is negative. Therefore, the fact that Alcubierre and Natário solutions are truncated may partly explain why they require negative energy even in subluminal regimes. Finally, for any spherically symmetric warp drive spacetime, since $\Lambda(0)=\Lambda(r\rightarrow\infty)=1$, coordinate $r$ describes the same physical length scale both inside the drive and at the asymptotic infinity. In other words, unlike the drives discussed in \cite{vandenBroeck1999}, spherically symmetric warp drives cannot contain objects which significantly exceed the sizes of warp drives, as measured by external observers.

We may further calculate the spatial component of the stress energy tensor. To do so, we assume that the material in region $\mathcal{D}_{\rm warp}$ is an isotropic fluid, i.e. $T_{rr}=P(r) \Lambda(r)$. This way, we obtain the pressure of the material which constitutes the warp drive:
\begin{equation}
    P(r)\Lambda(r) = \frac{1-\Lambda(r)+r N^\prime(r)/N(r)}{r^2}
    \label{eq:Sph3}
\end{equation}
By requiring that $P=0$ in the inner vacuum region $\mathcal{D}_{\rm in}$, and by remembering that $\Lambda(r)=1$ in that region, we can see that $N(r)$ is constant in $\mathcal{D}_{\rm in}$. In the external region $\mathcal{D}_{\rm out}$, from Equation~(\ref{eq:Sph3}), $N(r) = (\Lambda(r))^{-1} = 1-\frac{2}{r}\int_0^r 4\pi w(r^\prime) r^{\prime 2}{\rm d}r^{\prime}$.  This result also follows from Birkhoff's theorem, according to which spherically symmetric spacetimes must be static and asymptotically flat. In the vacuum region, such setup, by definition, corresponds to the Schwarzschild solution (assuming the central object is electrically neutral), which matches by the form the equation for $N(r)$. In this case, $\int_0^r 4\pi w(r^\prime) r^{\prime 2}{\rm d}r^{\prime}$ will have the effective meaning of enclosed mass.

We can now obtain $N(r)$ in region $\mathcal{D}_{\rm warp}$, and this way determine $N(r)$ in the inner region $\mathcal{D}_{\rm in}$. We do this by writing down the equation for continuity of the stress-energy tensor, $T^{\mu}_{~r;\mu}=0$, similarly to how it is done for Tolman–Oppenheimer–Volkoff (TOV) equation. As a result, we obtain:
\begin{equation}
    \frac{N^\prime(r)}{N(r)}=-\left(\frac{P+\rho}{2}\right)^{-1}P^\prime(r)
    \label{eq:Sph4}
\end{equation}
{ Firstly, we make a simplifying assumption that density $\rho$ is a function of only pressure $P$, as, for example, is the case for the polytropic equation of state. Then, by using the condition that $P=0$ at the inner and outer boundary of the warp region $\mathcal{D}_{\rm warp}$, we find that the integral of the right-hand side over $\mathcal{D}_{\rm warp}$ vanishes. As a result, we conclude that $N(r_{\rm in})=N(r_{\rm out})$, where $r_{\rm in}$ and $r_{\rm out}$ are the locations of the inner and outer boundary of the region $\mathcal{D}_{\rm warp}$.}

From the solution for $N(r)$ in the outside region $\mathcal{D}_{\rm out}$, we see that the time-time coefficient of the metric $N(r)$ at the asymptotic infinity can only be larger than inside the warped region. In other words, for a subluminal warp drive based on non-exotic matter, the time inside the drive can only pass more slowly than it does for a remote Minkowski observer comoving with the drive. For scale, an Earth-mass shell of $10$ meter radius will slow down the rate of time by a small fraction of $4\cdot 10^{-4}$. If we require that the time in the inner region goes faster than in the reference frame of the comoving observer, the material in the warp drive would have to contain negative energy. This is the case for the Alcubierre and Natário drives. In these drives, the time is passing more quickly in the inside region than it is in the reference frame of the remote comoving observer (due to the clocks in the inner region being synchronized to the remote observer at rest). This is another related possible reason why the Alcubierre drive requires negative energy even in the subluminal mode.


\subsection{Impossibility of superluminal spherically symmetric spacetimes}

The method from the previous section could potentially have been useful for analysing superluminal warp drives. However,  superluminal classes of warp drives (Classes II and III) cannot be spherically symmetric. This fact is closely related to the impossibility of superluminal spherical objects in special relativity, e.g. \cite{Fayngold2002}. In Section~\ref{sec:GenDrives}, we have introduced a Killing vector field ${\bf \xi}$, which defines a local frame of rest with respect to the drive. In superluminal warp drives, vector field ${\bf \xi}$ becomes spacelike whether inside or outside of the drive. In order both to maintain spherical symmetry and for the field ${\bf \xi}$ to remain orthogonal to the orbits of the ${\rm SO}(3)$ group, the spatial component of the field ${\bf \xi}$ can only be radial with respect to the drive. Therefore, at asymptotic infinity, comoving observers for such a spacetime would have to be moving radially towards or away from a spherically symmetric drive. In other words, warp drives cannot be both spherically symmetric and move superluminally.

Similarly, Class IV drives cannot be spherically symmetric. As follows from Equation~(\ref{eq:Sph2}), in order for the radial basis vector inside the drive to be timelike, one would need to have non-zero energy density present in the vacuum inner region $\mathcal{D}_{\rm in}$. Therefore, all spherically symmetric positive-energy warp drive solutions belong to Class I. These solutions are always subluminal, satisfy the energy conditions, and are devoid of causality paradoxes present in superluminal metrics. Overall, the subluminal spherically symmetric solution presented in this section is the first example of manifestly positive energy warp drive spacetimes.

\section{Axisymmetric warp drives with a general internal region}
\label{sec:Axi}

In this section, we explore the diversity of axisymmetric warp drive solutions. There are some limitations on possible axisymmetric solutions. Explicitly expressing a metric through a desired stress-energy tensor in a closed form is not possible for a general axisymmetric spacetime. Therefore, we leave generalizing the positive energy spherically symmetric warp drive solutions to the axisymmetric case to future studies. However, by constructing sufficiently broad classes of metrics, we show that interesting solutions are possible even within these less general classes. 

\subsection{Method for constructing axisymmetric warp drive solutions}
\label{sec:AxiClasses}

Below we introduce a method, by which one can construct new metrics for  a warp drive spacetime. For constructing new solutions, following the discussion in Section \ref{sec:SphGen}, we focus on choosing how the spacetime properties of the observers inside the inner region $\mathcal{O}_{\rm in}$ relate to those of remote observers $\mathcal{O}_{\rm out}$. This choice may put significant constraints on the energy content of the spacetime, and may potentially be exploited to find spacetimes with lower energy requirements. Additionally, having an explicit way to define the properties of spacetime inside the inner region $\mathcal{D}_{\rm in}$ allows us to explore and demonstrate the diversity of possible warp drive solutions.

As discussed in Section~\ref{sec:GenDrives}, any stationary warp drive spacetime may be associated with a coordinate system $x^\mu_{\rm co}$, adapted to the Killing vector ${\bf \xi}$ which defines the rest frame of the craft. We also adopt a global coordinate system $x^\mu$ which asymptotically, at infinity, approaches the coordinate system of the resting observer $\mathcal{O}_{\rm out}$ in region $\mathcal{D}_{\rm out}$. Since the two charts $x^\mu_{\rm co}$ and $x^\mu$ cover the whole spacetime and overlap, it is, in principle, possible to introduce a mapping from one to another in the regions $\mathcal{D}_{\rm in}$ and $\mathcal{D}_{\rm warp}$. The procedure for constructing an axisymmetric warp drive spacetime from this subclass is as follows:
\begin{enumerate}
    \item[1.] Choose a one-to-one mapping $x^{\mu}_{\rm co}(x^{\nu})$ between the coordinate system $x^{\mu}_{\rm co}$ adopted to the comoving observer inside the inner region $\mathcal{O}_{\rm in}$ and the coordinate system $x^{\nu}$ adopted to the external observer at rest $\mathcal{O}_{\rm out}$.
    \item[2.] Choose functions  $f_x (x^i_{\rm co})$, $f_y (x^i_{\rm co})$, $f_z (x^i_{\rm co})$, $f_t (x^i_{\rm co})$ which are equal to $1$ in region $\mathcal{D}_{\rm in}$ and are equal to $0$ in region $\mathcal{D}_{\rm out}$. These functions define the shape and size of the warp region $\mathcal{D}_{\rm warp}$ from the point of view of observer $\mathcal{O}_{\rm in}$.
    \item[3.] Formulate the metric of the spacetime as:
    \begin{equation}
    \label{eq:GenTrans}
    {\rm d}s^2=-c^2  ({\rm d}t(1-f_t)+f_t{\rm d} t_{\rm co})^2 + ({\rm d}x(1-f_x)+f_x {\rm d} x_{\rm co})^2+\qquad\
    \end{equation}
\[
 \qquad + ({\rm d}y(1-f_y)+f_y{\rm d} y_{\rm co})^2 + ({\rm d}z(1-f_z)+f_z{\rm d} z_{\rm co})^2,
\]
Express the metric in terms of { a common coordinate system $x^{\mu \prime\prime}$, by expressing $x^\mu_{\rm co}$ and $x^\nu$ explicitly through $x^{\mu \prime\prime}$}.
    \item[4.] Analyze the metric with the methods introduced in Section~\ref{sec:DriveProps}
\end{enumerate}

The procedure above is based on the idea of explicitly comparing the measurements of observers $\mathcal{O}_{\rm in}$ and $\mathcal{O}_{\rm out}$. Indeed, in region $\mathcal{D}_{\rm in}$ all the functions $f_{\eta}=1$, and the metric corresponds to the coordinate system of observer $\mathcal{O}_{\rm in}$ (${\rm d}s^2=-c^2 {\rm d} t_{\rm co}^2 + {\rm d} x_{\rm co}^2 + {\rm d} y_{\rm co}^2 + {\rm d} z_{\rm co}^2$). Similarly, asymptotically, in region $\mathcal{D}_{\rm out}$ all the functions $f_{\eta}=0$ and the metric corresponds to the coordinate system of the remote observer at rest $\mathcal{O}_{\rm out}$ (${\rm d}s^2=-c^2 {\rm d} t^2 + {\rm d} x^2 + {\rm d} y^2 + {\rm d} z^2$). Several studies in the literature, e.g. \cite{Loup2001}, have proposed modifications of the Alcubierre metric without explicitly considering the measurements of inner and outer observers. Such modifications may be erroneous and may reduce to coordinate transformations, as happens in \cite{Loup2001} case. We discuss these studies further in \ref{sec:AppA}.

The procedure developed here assumes that the observers $\mathcal{O}_{\rm in}$ and $\mathcal{O}_{\rm out}$ are timelike, and therefore leads to warp drive spacetimes of Class I in the subluminal regime, and of Class III in the superluminal regime. While these classes of spacetimes are arguably most interesting, the procedure may be easily generalized to cover the spacetimes of Class II and Class IV. We also comment that while the procedure ensures a relationship for the measurements of observers $\mathcal{O}_{\rm in}$ and $\mathcal{O}_{\rm out}$, one may also use it to obtain different coordinate representations and, potentially, different spacetimes satisfying the same relationships between the observers. Subsequently, one may choose, for example, the most interesting spacetimes of the class.

As a demonstration, we show how the above procedure may be used to construct the Alcubierre metric given by Equation~(\ref{sec:Intro}).

\begin{itemize}
    \item[1.] We choose the one-to-one mapping as follows:
    \begin{equation}
\left\{
\begin{array}{lll}
    {\rm d}t_{\rm co}&={\rm d}t\\
    {\rm d}x_{\rm co}&={\rm d}x - v_s{\rm d}t\\
    {\rm d}y_{\rm co}&={\rm d}y\\
    {\rm d}z_{\rm co}&={\rm d}z\\
    \end{array}
    \right.
\end{equation}
    \item[2.] We select $f_x=f$, where $f$ is given by Equation~(\ref{eq:ShFun0}), whereas functions $f_y$, $f_z$ and $f_t$ cancel out in Equation~(\ref{eq:GenTrans}) and will not contribute to the metric.
    \item[3.] We arrive at the metric ${\rm d}s^2=-c^2 {\rm d} t^2 + ({\rm d}x-f(r_s)v_s {\rm d}t)^2+{\rm d}y^2 + {\rm d}z^2$, as given by Equation~(\ref{eq:Alc}).
\end{itemize}

This example once again highlights the fact that the Alcubierre metric is based on a rather artificial identification of spacetime properties between the two observers $\mathcal{O}_{\rm in}$ and $\mathcal{O}_{\rm out}$, in particular increasing the rate of the time passing for the observer $\mathcal{O}_{\rm in}$. In the following sections, we explore other possible axisymmetric spacetimes and show that they have more appealing properties than the Alcubierre metric.

\subsection{Reducing $E_{\rm tot}$ by flattening the Alcubierre metrics}
\label{sec:AlcDeformed}

We start by considering the flattened Alcubierre metrics. The longitudinal extent is a simple property of the Alcubierre metric. However, it has not been studied in the literature, despite having several interesting properties.

The longitudinal extent has several peculiarities for the Alcubierre metric. Indeed, in Equation~(\ref{eq:Alc}), for sufficiently large distances $r_{s0}$ from the center of the warped region, function $f(r_s)$ vanishes in the Alcubierre solution. Therefore, the external observer $\mathcal{O}$ perceives the boundaries of the warped region $\mathcal{D}_{\rm warp}$ as a sphere of radius $r_{s0}$ moving with velocity $v_s$. If the velocity $v_s$ is subluminal due to Lorentz contraction, the boundary of the warped region $\mathcal{D}_{\rm warp}$ must be elongated for the comoving observers. Moreover, when the Alcubierre warp drive approaches the speed of light while at the same time preserving its energy content, given by Equation~(\ref{eq:AlcT}), the warped region must appear infinitely elongated. The divergent elongation of the solution in the comoving frame puts into question the possibility of accelerating Alcubierre metrics beyond the speed of light.

To analyze the solutions deformed along the axis of motion, we switch to more convenient cylindrical coordinates $(y,z)\rightarrow (\rho,\theta)$. For the transition regions described by Equation~(\ref{eq:Alc}), a more general non-spherically symmetric metric may be obtained by replacing $f(r_s)\equiv f(\sqrt{(x-x_s)^2+\rho^2}) \longrightarrow f(x-x_s, \rho)$. Notably\footnote{We believe, this form has not been found in the literature so far}, for the Alcubierre drive, recalculating the energy density with $f = f(x-x_s, \rho)$ simplifies the expression, compared to Equation~(\ref{eq:ShFun0}):
\begin{equation}
   w=-\frac{1}{8\pi}\frac{v_s^2}{4}\left(\frac{{\partial} f}{{\partial} \rho}\right)^2
\end{equation}

This form leads to three useful implications. The first implication is that, for a given velocity of the warp drive $v_s$, the most optimal way of reducing the total energy, as measured by Eulerian observers, is by flattening the shape of the warp drive. Indeed, as $\sqrt{-g}=\rho$ for the Alcubierre metric, the total energy 
$E=-\frac{v_s^2}{16}\int_{-\infty}^{\infty}{\rm d}x \int_0^{\infty}{\rho{\rm d}\rho}\left(\frac{{\partial} f}{{\partial} \rho}\right)^2$. And indeed, flattening the warp drive by a factor of $\alpha_X$ ($\alpha_X>1$) -- i.e., by replacing $f(x-x_s, \rho) \rightarrow f(\alpha_X(x-x_s), \rho)$, as may be shown through variable change $\alpha_X(x-x_s) \rightarrow x^\prime$ -- leads to the energy reduction by $E\rightarrow E/\alpha_X$. Similarly, elongating the drive (choosing $\alpha_X<1$) increases its energy requirements. Therefore, putting aside the fundamental issues related to negative energy of Alcubierre drives, an optimal implementation of such spacetimes would likely be flattened in shape.

The second implication of Equation~(\ref{eq:ShFun0}) is that the flattening of the warped region may be adjusted with velocity so that the drive preserves the same total energy. Indeed, setting $\alpha_X=1+v^2$ leads to $E\rightarrow E/(1+v^2)$, asymptotically removing the dependency of the total energy on velocity. As discussed earlier in Section~\ref{sec:Intro}, the fact that the total energy of the Alcubierre drive depends on velocity is problematic. This is because the energy and momentum conservation, applicable to asymptotically-flat spacetimes, implies that the Alcubierre drive must be changing its already very large energy (and mass) as it accelerates. Removing or softening the dependence of the total energy on velocity may, in principle, lead to more efficient ways of accelerating the drive to large velocities.

Finally, the third implication of Equation~(\ref{eq:ShFun0}) is that it allows one to construct superluminal solutions which satisfy the quantum inequalities given by Equation~(\ref{eq:EnCond}). Indeed, selecting a sufficiently large $\alpha_X$ allows one to reduce the thickness of the warp in $x$-direction down to nearly-Planck scale size (correspondingly, allowing only for extremely thin physical observers inside the warp bubble). As a result, the local curvature radius in Equation~(\ref{eq:EnCond}) may be arbitrarily small, thus satisfying the quantum inequalities given by Equation~(\ref{eq:EnCond}). At the same time, such superluminal drives still maintain a macroscopic size in $\rho$-direction perpendicular to the direction of motion and, more importantly, do not increase their densities due to contraction. Such solutions, may offer an exciting possibility of superluminal physical solutions. Perhaps, they may help probe the physics of superluminal motion and the problems associated with it, e.g. the violation of causality \cite{Krasnikov1998}, \cite{Everett1997}. However, they more likely probe the limits of applicability of the quantum inequalities~(\ref{eq:EnCond}), which are derived { in semiclassical gravity approximation. Indeed, superluminal motion violates the averaged null energy condition \cite{Visser2000}, and the latter does not have the same dependency on the dimensions of the bubble as the quantum inequalities, e.g. \cite{Graham2007}. As a result, even the extremely flattened version of the Alcubierre drive, as discussed here, does not satisfy the averaged null energy conditions.}

\subsection{Lorentz drive}

The Alcubierre metric is artificial in the sense that it forces the clocks and rulers of observers $\mathcal{O}_{\rm in}$ and $\mathcal{O}_{\rm out}$ to be synchronized. An arguably more natural choice would be to require that the observer $\mathcal{O}_{\rm in}$ should experience the same time dilation and space contraction as a Lorentz observer would experience when moving with velocity $v_s<c$. Since Lorentz transformations are defined for subluminal speeds, in this section, we consider subluminal warp drives of Class I. To construct the spacetime, we choose:
\begin{equation}
\label{eq:LorDrive}
    {\rm d}t_{\rm co}=\gamma({\rm d}t-\frac{v_s {\rm d}x}{c^2})
\end{equation}
\[
{\rm d}x_{\rm co}=\gamma ({\rm d}x-v_s {\rm d}t),
\]
where $\gamma$ is the Lorentz gamma-factor. Using these definitions to construct the warp drive metric with Equation~(\ref{eq:GenTrans}) leads to a diagonal metric:
\begin{equation}
{\rm d}s^2=-c^2 F^2{\rm d} t^2 + F^2{\rm d}x^2+{\rm d}\rho^2 + \rho^2{\rm d}\theta^2,
\end{equation}
where $F^2\equiv 1+2f(1-f)(\gamma -1)$. Moreover, any diagonalizable warp drive spacetimes are described by Equation~(\ref{eq:LorDrive}) (the diagonal form is retained if $\gamma$ is replaced by any constant).

The energy density for this metric is given by:
\begin{equation}
   w = -\frac{1}{8\pi}\frac{(\rho F^\prime_\rho)^\prime_\rho}{\rho F}
\end{equation}

For this metric, independent of the choice of function $f(x-x_s,\rho)$, the region $\mathcal{D}_{\rm warp}$ contains areas of both positive and negative energy density. This result is likely related to the fact that the spacetime is effectively flat beyond a certain radius, rather than asymptotically approaching the flat spacetime, similarly to the Schwarzschild solution. In the case of spherically symmetric solutions, as discussed in Section~\ref{sec:SphGen}, such truncation of the gravitational field also requires the warp region $\mathcal{D}_{\rm warp}$ to contain regions of positive and negative energy.

The Lorentz drive metric may be generalized to porduce a continuous limit to flat spacetime. Indeed, if region $\mathcal{D}_{\rm warp}$ were replaced by Minkowski spacetime, the whole spacetime would be Minkowski space. Therefore, one may introduce a continuum of solutions parametrized by a dimensionless parameter $\lambda\in[0,1]$, by using $F^2\equiv 1+2\lambda f(1-f)(\gamma -1)$. For $\lambda = 0$ the whole spacetime reduces to a flat spacetime, and for $\lambda = 1$ the Lorentz warp drive solution is recovered. The intermediate values of $\lambda$ allow for solutions with smaller energy requirements. To our knowledge, this is the first example of a family of warp drive solutions containing near-Minkowski metrics.

\subsection{An improved Van Den Broeck's drive}

The warp drive solution by \cite{vandenBroeck1999} is intended to reduce the energy requirements of warp drives by significantly expanding the volume of the Alcubierre drive inside radius $r_s$ while significantly decreasing the externally measured size of the craft. 
In this section, we realize the idea of \cite{vandenBroeck1999} in a simpler and, importantly, coordinate-independent form. We select the relation between clocks of the internal and the external observers to be the same as in the Alcubierre metric, ${\rm d}t_{\rm co}={\rm d}t$, but choose the internal spacetime to be expanded in $x$-direction: ${\rm d}x_{\rm co}=A({\rm d}x
-v_s{\rm d}t)$, ${\rm d}\rho_{\rm co}={\rm d}\rho$, ${\rm d}\theta_{\rm co}={\rm d}\theta$, where $A>1$. Substituting these relations into Equation~(\ref{eq:GenTrans}), we obtain the resulting spacetime:
\begin{equation}
    {\rm d}s^2=-c^2 {\rm d}t^2 + ({\rm d}x + f(A-1){\rm d}x-fAv_s{\rm d}t)^2 + {\rm d}\rho^2 + \rho^2{\rm d}\theta^2
\end{equation}
This spacetime leads to the energy density distribution of:
\begin{equation}
    w=-\frac{1}{8\pi}\left[\frac{v_s^2}{4}\frac{A^2f^{\prime 2}_{\rho}}{(1+(A-1)f)^2}+(A-1)\frac{(\rho f^\prime_\rho)^\prime_\rho}{(1+(A-1)f)}\right]
\end{equation}
The expression contains a velocity-independent component, which also leads to regions of positive and negative energy density. We see here that while stretching the space inside the warp drive does lead to higher energy density, the metric may be optimized to achieve the highest internal volume per unit energy needed to construct the spacetime.

\subsection{Warp drive metric with modified time}
In this section, we consider spacetimes, in which the clocks in the region $\mathcal{D}_{\rm in}$ run at a different rate compared to the clocks in the region $\mathcal{D}_{\rm out}$. This is done by letting ${\rm d}t_{\rm co}=A^{-1}{\rm d}t$, where $A>1$ corresponds to the clocks going slower in the region $\mathcal{D}_{\rm in}$. Substituting this relation into Equation~(\ref{eq:GenTrans}) and otherwise using the same definitions as in the Alcubierre solution, we obtain:
\begin{equation}
    {\rm d}s^2=-c^2 ((1-f){\rm d} t + A^{-1}f{\rm d} t)^2 + ({\rm d}x-f v_s {\rm d}t)^2 + {\rm d}\rho^2 + \rho^2{\rm d}\theta^2
    \label{eq:AlcV}
\end{equation}
The resulting energy density for the Eulerian observers is:
\begin{equation}
    w=-\frac{1}{8\pi}\frac{ (1-(A^{-1}-1)f)^2}{(1+(A^{-1}-1)f)^4}\cdot\frac{v_s^2}{4}\left(\frac{{\partial} f}{{\partial} \rho}\right)^2
    \label{eq:TChangeDrive}
\end{equation}
This demonstrates that the rate of clocks used in the original Alcubierre solution ($A=1$) in fact leads to the simplest possible expression for the total energy of the warp drive, other parameters being constant. From Equation~(\ref{eq:TChangeDrive}) it follows that slowing down the clocks inside the Alcubierre drive leads to higher amounts of energy density from the point of view of Eulerian observers.

\subsection{Spinning warp drive metric}

Finally, our method can be used to construct relatively complex relations between the inner and outer regions. For example, the following metric corresponds to the two Minkowski regions $\mathcal{D}_{\rm in}$ and $\mathcal{D}_{\rm out}$ spinning with respect to each other. In other words, the observers at rest in region $\mathcal{D}_{\rm in}$ will be rotating with respect to observers in region $\mathcal{D}_{\rm out}$ without experiencing any centrifugal or Coriolis forces typical to rotating systems. Such settings are impossible to realize in the absence of the transition region $\mathcal{D}_{\rm warp}$. The settings may be achieved by relating ${\rm d}\theta_{\rm co}={\rm d}\theta - \omega_s {\rm d}t$, in addition to the definition used for the Alcubierre drive, where $\omega_s$ is a constant parameter defining the angular velocity. Substituting these relations into Equation~(\ref{eq:GenTrans}) leads to the following metric:
\begin{equation}
    {\rm d}s^2=-c^2 {\rm d} t^2 + ({\rm d}x-v_s f {\rm d}t)^2+{\rm d}\rho^2 + (\rho{\rm d}\theta-f\rho \omega_s {\rm d}t)^2,
    \label{eq:AlcAlt}
\end{equation}
The energy density for Eulerian observers then is:
\begin{equation}
    w=-\frac{1}{8\pi}\left[\frac{v_s^2}{4}\left(\frac{{\partial} f}{{\partial} \rho}\right)^2 + \frac{\rho^2\omega_s^2}{4}\left(\left(\frac{{\partial} f}{{\partial} \rho}\right)^2 + \left(\frac{{\partial} f}{{\partial} x}\right)^2\right)\right]
\end{equation}
We see that the energy density is negatively defined independent of the spin or velocity of the inner region. Additionally, the contributions from the velocity in $x$-direction and the velocity of angular rotation have a similar form. The main interest for such spacetimes may be that they offer a stationary, dissipationless way of storing energy or angular momentum.








\section{Discussion}
\label{sec:Disc}

\subsection{What is a warp drive?}

One of the main conclusions of our study is that warp drives are simpler and much less mysterious objects than the broader literature has suggested when citing \cite{Alcubierre1994}. Warp drives are inertially moving shells of positive or negative energy material which enclose a `passenger' region with a flat metric. The main feature distinguishing warp drives from trivial inertially moving low-mass shells is that the large amount of energy contained in the warp shell allows one to modify the state of spacetime inside it. In particular, as shown in Sections~\ref{sec:SphGen} and~\ref{sec:Axi}, the time in the inner region may go faster or slower than it would go without the shell. Similarly, the spatial volume may be stretched, compressed, or even be rotating compared to its normal state. Further, more complex, modifications are likewise possible.

Warp drives can move superluminally only in the same sense as any ordinary inertial mass, test mass, or any other object. Namely, there is no known way of accelerating regular material beyond the speed of light. However, one may postulate a test particle which moves faster than light in relativity, in which case it may continue moving inertially. In the same way, as warp drives are shells of material, there is no known way of accelerating a warp drive beyond the speed of light. However, one may also postulate the warp drive shell to be in superluminal motion, just like the hypothetical test particles, and the shell-like object will continue moving in the same fashion. In this sense, superluminal warp drives are at least as hypothetically possible as any other superluminal objects.

An interesting feature of warp drives, from the theoretical point of view, is that the modifications of the spacetime in the internal region may be sufficiently strong so as to allow superluminal objects to move subluminally or vice versa. The different possibilities are embodied in four different classes of warp drives, presented in Section~\ref{sec:GenDrives}. In particular, there are solutions possible (Class IV) wherein a shell of subluminal material may contain regions where no subluminal material can remain at rest no matter what velocity it has. In this aspect, the shells of these hypothetical Class IV warp drives would share properties with black holes. Further, such a spacetime could effectively stop and contain inside a hypothetical superluminal test-mass object, or make it move slower than the speed of light. 

Similarly, one may hypothesize a spherical shell moving faster than the speed of light, but which contains an inner region where no (hypothetical) superluminal objects can be at rest relative to the shell (Class III). Or, in another interpretation, subluminal normal objects inside such shell could be at rest with respect to them, despite their motion being superluminal. This class of warp drives, which also includes the Alcubierre drive, however, remains entirely hypothetical, just in the same way as no other object can be set to move superluminally. { Using the methods in Section~\ref{sec:GenDrives}, one can examine whether any other particular spacetime belongs to this class of warp drives.}

\subsection{Constructing warp drives}

Warp drives, being inertially moving shells of normal or exotic material, do not have any natural way of changing their velocities. They are just like any other types of inertially moving objects. Similarly, just like for any other massive objects, achieving a certain velocity for a warp drive requires an externally applied force or, more practically, some form of propulsion. Propulsion may be realized, for example, by an interaction with a bosonic field, or regular gaseous or plasma material.

Whatever is the acceleration mechanism, it must obey the conservation of 4-momentum. This is because all warp drive spacetimes are asymptotically-flat. An unfortunate error, introduced in \cite{Alcubierre1994}, was to postulate the velocity in Equation~(\ref{eq:Alc}) to be time-variable. An Alcubierre spacetime with time-variable velocity also changes its energy and momentum with time, and, this way, such a construction violates energy conservation. More technically, the metric given by Equation~(\ref{eq:Alc}) does not satisfy the continuity equations, unless additional dynamical fields are implicitly introduced to compensate for that. In view of this, no metric which describes an accelerating warp drive solution has so far been presented in the literature.

A more subtle point is that the Alcubierre and Natário drives, as well as the spacetimes constructed in Section~\ref{sec:Axi}, represent classes of different objects parametrized by velocity $v$ rather than the same object changing its velocity $v$. In particular, different warp drive solutions with different values of $v$ have different mass, different energy content, and often different shapes in their reference frame of rest. Any realistic object should at least conserve its ADM-mass in the subluminal regime and its analogue in the superluminal regime. A natural way of constructing such spacetimes is by defining them explicitly in their frame of rest, as discussed in Section~\ref{sec:GenDrives} and implemented for spherically symmetric drives in Section~\ref{sec:SphGen}. Implementing metrics for such accelerating objects and more general axisymmetric objects which preserve their shape and mass in the comoving frame as they change their velocity remains a subject for future studies. As a simple compromise, one may also adjust the shape of the warp drive with velocity so as to conserve the mass of the drive, as suggested in Section~\ref{sec:AxiClasses}. Finally, among all classes of subluminal warp drive solutions, the particularly interesting ones in the practical sense are those classes which contain a continuous set of solutions ranging from trivial to highly curved.

At least in the subluminal case, warp drive spacetimes may be constructed by using purely positive energy density, as presented in Section~\ref{sec:SphGen} for the spherically symmetric case. They can likewise be constructed using purely negative energy density, as is the case for the Alcubierre solution, or constructed using both positive and negative energy density. In Section~\ref{sec:SphGen} we showed, for the first time, that the only type of modification to the internal spacetime that is achievable with purely positive energy for spherically symmetric warp drives is slowing down the rate of time inside the craft.

In Section~\ref{sec:Axi} we demonstrated that, by using both positive and negative energy density, one may achieve a variety of modifications for the spacetime inside more general axisymmetric subluminal warp drives. The range of all the possible modifications achievable with purely positive energy in the general axisymmetric case, and whether the class of axisymmetric spacetimes in Section~\ref{sec:Axi} may lead to purely positive energy metrics, remain important open questions in the field.  Similarly, in Section~\ref{sec:GenDrives} we have provided a new argument why superluminal warp drive solutions may always violate weak energy conditions resulting in their negative energy density requirements. While this is an established fact \cite{Olum1998,Visser2000}, a strict independent proof of this based on our argument is another important avenue for future studies. Our conclusions do not support the recent claim in \cite{Lentz2020} of superluminal purely positive energy warp drive solutions, which merits further investigation.

\subsection{Optimal members in the class}

An interesting question remaining is: assuming that a practical realization of a warp drive spacetime is possible, what would the optimized versions look like? As we showed in Section~\ref{sec:AlcDeformed}, a more optimal implementation of the Alcubierre warp drives would be flattened in shape, since such shapes are more efficient in terms of energy requirements. In particular, flattening the shape by a factor of $10$ would lead to proportionally smaller energy requirements. This conclusion likely holds for the, more physical, purely positive energy subluminal warp drives as well. Curiously, as we discussed in Section~\ref{sec:AxiClasses}, extreme flattening of Alcubierre drives may allow for superluminal solutions which satisfy quantum inequalities, without reaching extreme energy densities.

For the Alcubierre solution, one may similarly optimize the energy requirement by finding the most suitable shape function $\bar{f}(r_s)$. The originally proposed function given by Equation~(\ref{eq:ShFun0}), is not optimized and was originally chosen solely for demonstration. By applying the variational method to the expression for the total energy of the Alcubierre drive, we find that the shape function optimizing the energy is given by $\bar{f}(r_s)=\min(\frac{r_0}{r_s},1)$, where $r_0$ is a free parameter determining the inner size of the region $\mathcal{D}_{\rm warp}$. Using this slower-decreasing shape reduces the energy requirement for a similarly sized Alcubierre drive by about a factor of three. The physical reason for it, as we discuss in Section~\ref{sec:SphGen}, is perhaps related to the fact that truncating the gravitational fields of a warp drive, as done in the Alcubierre solution, may increase the (absolute) amount of necessary negative energy compared to the more slowly falling off solutions. We provide the details of this derivation in \ref{sec:AppA3}, and also remark that optimizing the shape or matter distribution in region $\mathcal{D}_{\rm warp}$ can be equally well performed for all the other warp drive solutions.

Given the wide range of possible states of spacetime achievable inside a warp drive, it is also possible to imagine more complex and instrumental optimizations. For example, one may speculate that it is possible, at least in principle, to form a region inside a subluminal warp drive which is similar to ergospheres of spinning black holes. In this case, such a region would be used as an efficient energy storage. The energy could then potentially be extracted through a Penrose process applied to the propellant of the craft, when passing through the ergoregion. Similarly to the Penrose process for spinning black holes, the extracted energy would likely be coming from the rotation of some regions of the spacetime.

Finally, since all warp drive objects require propulsion in order to accelerate, any practical implementation of such objects would have to be asymmetric in shape, since the back part would have to accommodate a propellant exhaust system. One may further hypothesise on setups, wherein black hole-like regions of the spacetime may be used to produce accretion power. Accretion of material onto black holes is known to be a few tens of times more efficient at extracting rest-mass energy in the form of electromagnetic radiation from the material than nuclear burning \cite{Frank2002}. Such a process could potentially provide both a source of energy and a source of propulsion.

\subsection{Towards physical warp drives}

In Section~\ref{sec:SphGen}, we demonstrated that it is possible to construct non-trivial warp drive solutions with purely positive energy. In other words, at least in principle, one can construct objects of progressively larger masses and with progressively more salient modifications to the internal spacetime. While the mass requirements needed for such modifications are still enormous at present, our work suggests a method of constructing such objects based on fully understood laws of physics.

The most promising way of practically probing such spacetimes is through laboratory experiments -- most importantly -- through analogue gravity experiments, e.g. \cite{Barcelo2005}. Another important avenue of exploring such spacetimes, especially the accelerating solutions, is through numerical relativity. Such experiments may bring a better understanding of purely positive energy drives, and negative energy solutions, as well as the possibility of accelerating objects superluminally.

Since the introduction of \cite{Alcubierre1994}, much theoretical effort has been put into unveiling the unphysical nature of the Alcubierre solution. Our work shows that there is a variety of warp solutions, each with properties often much more physical and interesting than the originally proposed spacetime. Through this, we suggest there is a need for broader theoretical and experimental investigation to uncover the full diversity and properties of physical warp drives.

\section*{Acknowledgements}
We would like to thank Parsa Ghorbani, Lorenzo Pieri, Adam Lewis, Philip Chang, Hrant Gharibyan and Stefano Liberati for their insightful comments and helpful discussions at different stages of this work.

\newpage 

\appendix
\section{Existing warp drive solutions}
\label{sec:AppA}
The previously mentioned Alcubierre and Natário metrics are distinct from each other and satisfy the definition of warp drive metrics. In this section, we list the metrics present in the literature that claim or intend to describe new warp drive metrics. We further show that these metrics reduce to the Alcubierre and Natário solutions. We also comment that  the \cite{Lentz2020} study likely forms a new class of warp drive spacetimes, though it does not provide means for reproducing itself.

\subsection{Loup metric}

The non-refereed study by \cite{Loup2001}, also discussed in the \cite{Alcubierre2017} review, aims to reduce the energy requirements of the \cite{Alcubierre1994} drive by introducing a lapse function, which modifies the time-components of the metric as a function of spatial coordinates. However, their metric is equivalent to the Alcubierre metric.

Indeed, their defining Equations~(7)--(9), expressed in our notation, read:
\begin{equation}
    {\rm d}s^2=- A^2(r_s) c^2 {\rm d} t^2 + ({\rm d}x-f(r_s)v_s {\rm d}t)^2+{\rm d}y^2 + {\rm d}z^2,
    \label{eq:Loup}
\end{equation}
where $A(r_s)\geq1$ is the lapse function, defined to be equal to unity at $r_s=0$ and, asymptotically, at $r_s\rightarrow \infty$. However, by transforming to a new time coordinate $A(r_s){\rm d} t \rightarrow {\rm d}\bar{t}$, and checking that the coordinate transformation has a finite non-vanishing Jacobian and is, therefore, well-defined, we arrive at:
\begin{equation}
    {\rm d}s^2=- c^2 {\rm d} \bar{t}^2 + ({\rm d}x-\bar{f}(r_s) v_s {\rm d}\bar{t})^2+{\rm d}y^2 + {\rm d}z^2,
    \label{eq:Loup2}
\end{equation}
where  $\bar{f}(r_s)\equiv \frac{f(r_s)}{A(r_s)}$. Therefore, the Loup metric is that of the Alcubierre drive.

The shape function, $\bar{f}(r_s)$, as in the Alcubierre solution, is equal to $1$ at $r_s=0$ and asymptotically approaches $0$ at infinity. Unlike in the Alcubierre solution, the shape function now decreases non-monotonically, which corresponds to redistributing the energy density, according to Equation~(\ref{eq:AlcT}). Assuming the volume of the inner flat region is preserved, such a shape function only increases the total energy required by the solution.

\subsection{Van Den Broeck metric}

The study by \cite{vandenBroeck1999} provided a metric, which intended to significantly reduce the energy requirements compared to the Alcubierre solution. Such a reduction was made by reducing the outer surface area of the warped region $\mathcal{D}_{\rm warp}$ and through expanding the volume in the interior region $\mathcal{D}_{\rm in}$. 

The \cite{vandenBroeck1999} metric is given as:
\begin{equation}
{\rm d}s^2= -c^2 {\rm d} t^2 + B^2(r_s)\left(({\rm d}x-f(r_s)v_s {\rm d}t)^2+{\rm d}y^2 + {\rm d}z^2\right),
\label{eq:vandenBroeck}
\end{equation}
where $B(r_s)\geq 1$ is a monotonically decreasing function, taking large values $B\gg 1$ at $r_s=0$ and asymptotically decreasing to unity. By applying a coordinate transformation defined by ${\rm d}{\bar{x}} = {\rm d}{x}B(r_s)$, ${\rm d}{\bar{y}} = {\rm d}{y}B(r_s)$, ${\rm d}{\bar{z}} = {\rm d}{z}B(r_s)$ and $\bar{t}=t$ and ensuring that the transformation is well-defined, by checking that the Jacobian of the transformation is finite and non-vanishing, the metric transforms to:
\begin{equation}
{\rm d}s^2= -c^2 {\rm d} \bar{t}^2 + \left({\rm d}\bar{x}-\frac{f(r_s)B(r_s)}{B(0)}(v_s B(0)){\rm d}\bar{t}\right)^2+{\rm d}\bar{y}^2 + {\rm d}\bar{z}^2
\label{eq:vandenBroeck2}
\end{equation}
At the center of the warped region $r_s=0$ or, equivalently, $x=v_s t$. In the new coordinates this corresponds to $\bar{x}=B(0)v_s t$. Therefore, in the new coordinates, the object moves with velocity $\bar{v}_s\equiv B(0)v_s$. Finally, introducing a new shape function $\bar{f}(\bar{r}_s)=\frac{f(r_s)B(r_s)}{B(0)}$, we arrive again at the Alcubierre metric:
\begin{equation}
{\rm d}s^2= -c^2 {\rm d} \bar{t}^2 + ({\rm d}\bar{x}-\bar{f}(\bar{r}_s)\bar{v}_s{\rm d}\bar{t})^2+{\rm d}\bar{y}^2 + {\rm d}\bar{z}^2
    \label{eq:vandenBroeck3}
\end{equation}
The modified shape function $\bar{f}(\bar{r}_s)$ satisfies the condition $\bar{f}(0)=1$ and decreases to zero asymptotically at large values of $\bar{r}_s$. As for the Alcubierre solution, ${\rm d}\bar{x}-\bar{v}_s{\rm d}\bar{t}$, ${\rm d}\bar{y}$, ${\rm d}\bar{z}$, ${\rm d} \bar{t}$ correspond to the coordinates adapted to a resting observer inside the inner region $\mathcal{D}_{\rm in}$. Therefore, the physical size of the inner region of the metric, as measured by the inner observer, is given by the region where function $\bar{f}(\bar{r}_s)$ is close to unity. In the Van Den Broeck example, function $B(r_s)$ is sharp-peaked at the center, $B(0)=B(\tilde{R})\approx 10^{17}$, and $B(\tilde{R}+\Delta)=1$, where $\tilde{R}=\Delta = 10^{-15}\,{\rm m}$. In the coordinates of the internal observer, $\mathcal{O}_{\rm in}$, the inner region is limited to $r_s=\tilde{R}$, or $\bar{r}_s=B(0) r_s= 100\,{\rm m}$, which corresponds to the physical size of the inner region. The location, where function $B(r_s)$ decreases to $1$, corresponds to $\bar{r}_s\approx 200\,{\rm m}$, at which point $\bar{f}(\bar{r}_s)=10^{-17}$, i.e. is nearly vanishing. In the Van Den Broeck example, function $\bar{f}(\bar{r}_s)$ should subsequently decrease further to zero, between distances corresponding to $r_s=R$ and $r_s=R+10^2v_sL_P$, i.e. in a thin region $100 v_s$ Planck scales thick. Since, at these distances, $B(\bar{r}_s)=1$, the interval of $\bar{r}_s$, at which the function $\bar{f}(\bar{r}_s)$ decreases from $10^{-17}$ to $0$, is also $100 v_s$ Planck scales thick.

Therefore, in summary, the Van Den Broeck solution is equivalent to the Alcubierre solution. The shape function $f(\bar{r}_s)$ in their study is chosen to decrease to nearly zero within a volume comparable to the inner volume of the drive (Region 1). Subsequently, further out, the function $f(\bar{r}_s)$ decreases to exactly zero (Region 2). Since in the second outer region (Region 2), function  $f(\bar{r}_s)$ decreases from a very small initial value to zero, it is expected that Region 2 should correspond to a small total energy and satisfy the quantum inequalities due to its near-zero thickness. As follows from our derivation, the total energy in the inner Region 1, and of the Van Den Broeck metric as a whole, is comparable to that of the standard Alcubierre solution of similar dimensions. Our derivation suggests that the total energy should be proportional to $B(0)^2$, through the $\bar{v}_s^2$ term. The absence in the van den Broeck expression for Region 1 of dependence on $B(0)$ or on the velocity at all, potentially explains why the energies they obtain for that region are small.

\subsection{Optimizing the energy requirements of the Alcubierre solution}
\label{sec:AppA3}

This section provides details on how one may optimize a warp drive solution in terms of its energy requirements. As an example, we choose the well-known Alcubierre solution. The total energy measured by Eulerian observers on a hypersurface of constant $t$ for this spacetime is given, as follows from Section~\ref{sec:DriveProps}, by $E=\int_{\mathcal{D}_{\rm warp}}(-g^{00})^{-1}T^{00}\sqrt{-g}{\rm d}^3 x^i_{\rm out}$. For the Alcubierre solution~(\ref{eq:Alc}), one may verify that the contravariant time-time component of the metric tensor and the metric determinant are equal to minus unity, i.e. $g^{00}=g=-1$. In this case, by using Equation~(\ref{eq:AlcT}) for the value of energy density $T^{00}$, the total energy is obtained from a simple expression:
\begin{equation}
    E = -\int_{\mathcal{D}_{\rm warp}} \frac{1}{8\pi}\frac{\rho^2 v_s^2}{4r_s^2}\left(\frac{{\rm d} f}{{\rm d} r_s}\right)^2 {\rm d}^3 x 
\end{equation}
Function $f(r_s)$ in this equation defines the location of the wall of the warp bubble and is given by Equation~(\ref{eq:ShFun0}). As we discuss in the main text, the specific form of the function was chosen in \cite{Alcubierre1994} rather arbitrarily in order to satisfy the requirement that $f(r_s)=1$ for $r_s\rightarrow0$ and $f(r_s)=0$ for $r_s\rightarrow\infty$. Therefore, one may search for other functions satisfying the same constraints and leading to some further desired properties, for example, an optimised energy.

To formulate a variational problem, we switch to spherical coordinates $(r_s,\theta,\varphi)$ centered at $x=x_s(t)$, $y=z=0$, with the pole aligned with the direction of motion. In these coordinates, $\rho = r_s \sin\theta$, and we get:
\begin{equation}
    E = -\frac{v_s^2}{16}\int_{r_0}^\infty{\rm d}r_s\int_0^{\pi}{\rm d}\theta r_s^2\sin^3\theta\left(\frac{{\rm d} f}{{\rm d} r_s}\right)^2 = -\frac{v_s^2}{12}\int_{r_0}^\infty{\rm d}r_s r_s^2\left(\frac{{\rm d} f}{{\rm d} r_s}\right)^2
\end{equation}
The Lagrangian for this system is $\mathcal{L}=r_s^2 f^{\prime 2}_{r_s}$, and therefore the Euler-Lagrange equation for the function $f(r_s)$ which optimizes the energy reads:
\begin{equation}
    \frac{\partial}{\partial r_s} \frac{\partial \mathcal{L}}{\partial f^{\prime}_{r_s}} = 2\frac{\partial}{\partial r_s} (r_s^2 f^{\prime}_{r_s}) = \frac{\partial \mathcal{L}}{\partial f} =  0
\end{equation}
The solution of this equation is $\bar{f}(r_s) = C + \frac{D}{r_s}$. The requirement that $f(r_s)=0$ at $r_s\rightarrow\infty$ can be satisfied by setting $C=0$, while the requirement that $f(r_s)=1$ for $r_s\rightarrow0$ may be satisfied by setting $D=r_0$. Since the variational problem was solved for $r>r_0$, the optimal solution in the whole space is given by $\bar{f}(r_s)=\min(\frac{r_0}{r},1)$. 

One may verify numerically that this choice of $\bar{f}$ decreases the needed (absolute value of) negative energy of the Alcubierre solution by about a factor of 3. Intuitively, the decrease may be understood because a smooth fall-off is more natural for gravitating bodies than a sharp-exponential cut in the metric introduced by Equation~(\ref{eq:ShFun0}). Repeating the derivation but assuming that $f$ is a general axisymmetric function of both $\rho$ and $x-x_s$ shows that in this case the energy is optimized by an infinitely thin shape, similar to the conclusions we obtained in Section~\ref{sec:Axi}. We conclude this section by mentioning that one can apply a similar method to other warp drive spacetimes (or even classes of warp drive spacetimes) to optimize their properties such as the total energy.


    

\section*{Bibliography}

\bibliographystyle{jphysicsB}
\bibliography{Lits}

\end{document}